\documentclass[aps,prd,floatfix,amsmath,amssymb,preprint,eqsecnum,nofootinbib]{revtex4}
\usepackage{graphicx}
\usepackage{dcolumn}
\usepackage{bm}
\usepackage{epstopdf}
\usepackage{amsmath, amsfonts, amssymb}
\usepackage{pstricks}
\usepackage{amsxtra}
\usepackage{amsthm}

\newcommand{\beq}{\begin{equation}}
\newcommand{\eeq}{\end{equation}}
\newcommand{\ba}{\begin{array}{ccc}}
\newcommand{\ea}{\end{array}}
\newcommand{\nn}{\nonumber \\}

\def\bea{\begin{eqnarray}}
\def\eea{\end{eqnarray}}

\usepackage{setspace} 
\begin{document}
\title{A model of a Fermi liquid using gauge-gravity duality}
\author{Subir Sachdev}
\affiliation{Department of Physics, Harvard University, Cambridge MA
02138}

\date{July 26, 2011 \\
\vspace{1.6in}}
\begin{abstract}
We use gauge-gravity duality to model the crossover from a conformal critical point to a confining Fermi liquid, driven by a change
in fermion density. The short-distance conformal physics is represented by an anti-de Sitter geometry, which terminates
into a confining state along the emergent spatial direction. The Luttinger relation, relating the area enclosed by the Fermi surfaces
to the fermion density, is shown to follow from Gauss's Law for the bulk electric field. 
We argue that all low energy modes are consistent with Landau's Fermi liquid theory. An explicit solution is obtained for the Fermi liquid for the case of hard-wall boundary conditions in the infrared.
\end{abstract}

\maketitle

\section{Introduction}
\label{sec:intro}

Much recent work \cite{nernst,sslee0,denef0,hong0,zaanen1,hong1,denef,faulkner,polchinski,gubserrocha,hong2,kiritsis,sean1,sean2,sean3,seanr,larus1,larus2,eric,kachru2,kachru3,kachru4,trivedi,zaanen2,ssffl,mcphys,liza,leiden,hong4,pp,waldram,yarom}
has described the compressible, non-superfluid states of quantum matter by
the methods of gauge-gravity duality. A variety of ``non-Fermi liquid'' and claimed ``Fermi liquid" phases appear to have been obtained,
but their correspondence with the known strongly-correlated phases of condensed matter models with short-range interactions
remains uncertain \cite{ssffl}. 
The recent models of Refs.~\onlinecite{sean3,hong4} are perhaps the closest to realizing a conventional
Fermi liquid state, in that they have $N$ Fermi surfaces (with $N \rightarrow \infty$) enclosing areas which add up to the total fermion density, as required by the Luttinger relation. However, there are low energy excitations
associated with infrared ``Lifshitz'' or related geometries, and these do not correspond to known low energy modes
of a Fermi liquid. Moreover, there are also low energy excitations at small momenta associated with Fermi
surfaces with vanishing Fermi momenta, and the extra low energy modes are possibly linked to each other.

We will begin with conformally-invariant quantum critical point in 2+1 spacetime dimensions
(possible recent examples are in Ref.~\onlinecite{tasi,semenoff}). This has a familiar gravitational
representation in a AdS$_4$ geometry. Then we apply a chemical potential $\mu$, and assume the system ultimately
crosses over to a confining and gapless Fermi liquid state with a small number of Fermi surfaces (including the case with just a single Fermi surface).
A Fermi liquid is a confining state because its {\em only\/} low energy
excitations are Fermi surface quasiparticles which do not carry charges of the gauge-field of the boundary conformal field theory \cite{liza}.
Field-theoretic models which can exhibit such a crossover were
discussed in Ref.~\onlinecite{liza}. This crossover to confinement will be manifested in a deviation
from the AdS$_4$ geometry in the infrared. 

As an aside, we note that at non-zero density, it is possible for a state to be both confining and gapless even without
the Goldstone bosons of a broken continuous symmetry,
as in a Fermi liquid. On the other hand, at zero density, the usual assumptions of particle physicists apply,
and the only possible gapless excitations in a confining state are Goldstone bosons.

Initial works \cite{nernst,sslee0,hong0,zaanen1,hong1,denef,faulkner,hong2} 
accounted for the chemical potential only via a Maxwell term
for the bulk gauge field, and this led to a Reissner-Nordstr\"om black brane, with a AdS$_2 \times R^2$
near-horizon infrared geometry at zero temperature. Subsequent 
works \cite{polchinski,sean1,sean2,sean3,seanr,larus1,larus2,eric,zaanen2,leiden,hong4,pp} included back-reaction on the metric 
from the matter in a Thomas-Fermi approximation, leading to a Lifshitz geometry in the infrared.
However, it is clear that neither of these geometries can represent a true confining state (such as a Fermi liquid \cite{liza,hong4}), 
given the many emergent collective excitations in the infrared. 

At zero chemical potential, the crossover from conformality to confinement has previously been understood
via a termination of the anti-de Sitter geometry \cite{witten,meyer} into a ``AdS soliton'' geometry.
Such a geometry has also been used more recently \cite{tadashi,gary}, for confining states at non-zero chemical potential in the presence of
scalar fields. Here we will examine a terminated AdS$_4$ geometry in the presence of a non-zero fermion density.
For numerical simplicity, we will use a simple `hard-wall' termination \cite{ssqcd,igor}, but we expect similar results
to apply to other confining boundary conditions in the infrared \cite{soft1,soft2,soft3,soft4}.
It is important to note that while all excitations are gapped in a terminated geometry at zero chemical potential, this is no longer
true once the chemical potential is taken large enough to obtain a non-zero density. Then there are gapless excitations across the Fermi surface.
Indeed, these are the only gapless excitations, and this is the key to realizing a Fermi liquid.

Another important feature of our analysis will be an exact treatment of the quantum mechanics of the fermions, avoiding
the Thomas-Fermi limit. Our initial mean field theory in Section~\ref{sec:mft} 
will treat the bulk metric and gauge field classically, but we will discuss consequences
of their quantum fluctuations in Section~\ref{sec:beyond}. We describe cases where the boundary
theory has a small number of Fermi surfaces, and focus on the case with a single Fermi surface. The previous
Thomas-Fermi analyses effectively had a continuum of an infinite number of Fermi surfaces; equivalently, their fermion wavefunctions
were delta-functions along the emergent spatial direction. The necessity of moving away from this limit has been
previously noted \cite{seanr,zaanen2,leiden}.

The presence of quantum fermions implies that 
their contribution to the average local currents and stresses are {\em not\/} local functions of the background fields: they depend
on the fermion eigenfunctions, which are controlled by the full spatial dependence of the background fields. Consequently, 
the problem of determination
of the background fields cannot be reduced to set of coupled ordinary differential equations. In this respect, we differ from 
other recent work \cite{zaanen2,leiden} which approximates fermion bilinears with a zero frequency contribution.
Our formalism focuses on determining the bulk eigenstates of the fermions, in the absence of external sources. 
Knowledge of these allows subsequent 
computation via spectral representations of the bulk-to-bulk Green's functions, and also of the boundary Green's functions, which represent the responses to sources. 

\section{Mean field equations and their solution}
\label{sec:mft}

We begin by setting up the framework for the simplest discussion of a Fermi liquid state.

We will model a Fermi liquid in 2+1 dimensions by a theory 
of `quantum electrodynamics' in a 4-dimensional spacetime with metric $g$.
The emergent spatial-coordinate is $z$, and the metric is asymptotically AdS$_4$ near the boundary 
$z \rightarrow 0$. We will work in Euclidean time, and assume the metric only
has diagonal components non-zero.
Apart from the terms which depend only upon $g$, the action is
\beq
S = \int d^4 x \sqrt{g} \left[   \frac{1}{4e^2} F_{ab}F^{ab}
+ i \left( \overline{\psi} \Gamma^M D_M \psi + m \overline{\psi} \psi \right) \right]. \label{qed}
\eeq
Here $F_{ab}$ is the gauge flux of a U(1) gauge field $A_\mu$, which is the bulk representation of the 
globally conserved U(1) charge $\mathcal{Q}$ in the boundary theory. The bulk theory also has
Dirac fermions with U(1) charge $q$ and mass $m>0$, and we will use the convention followed in
Refs.~\onlinecite{hong1,trivedi} for the Dirac matrices.

In the present section we treat the gauge field $A_\mu$ classically, while the fermions $\psi$ will be treated
exactly. In the absence of sources, the only non-zero component of the gauge field is its time-component (the electrochemical
potential)
$A_t  = i \Phi (z)$, and we can write the free energy density as
\beq
\mathcal{F} = - \frac{1}{2 e^2} \int dz \sqrt{g} g^{zz} g^{tt} \left( \frac{d \Phi}{d z} \right)^2 - \frac{T}{V} \mbox{Tr} \ln
\left[ \Gamma . \hat{D} + m \right] \label{free}
\eeq
where $T$ is the temperature, and $V$ is the spatial volume of the boundary theory. The determinant in Eq.~(\ref{free})
is to be evaluated as described in Ref.~\onlinecite{denef}. Now our task is simply to solve the equations
\beq 
\frac{\delta \mathcal{F}}{\delta \Phi (z)} = 0 \quad , \quad \Phi( z \rightarrow 0 ) = \mu, \label{eom}
\eeq
for $\Phi (z)$, and describe the properties of the resulting state of matter. In particular, the charge density of the boundary
theory, $\langle \mathcal{Q} \rangle$ is given by
\beq
\langle \mathcal{Q} \rangle = - \frac{ \partial \mathcal{F}}{\partial \mu}. \label{defQ}
\eeq
Eqns.~(\ref{eom}) and (\ref{defQ}) imply that $\langle \mathcal{Q} \rangle$ can be determined from Eq.~(\ref{free}) by Hamilton-Jacobi
theory to be
\beq
\langle \mathcal{Q} \rangle = - \frac{1}{e^2} \lim_{z \rightarrow 0} \frac{d}{dz} \Bigl[ \sqrt{g} g^{zz} g^{tt} \Phi (z) \Bigr] ; \label{surface}
\eeq
The right-hand-side is just the electric field along the $z$ direction at the boundary, and this is Gauss's law for
the ``surface'' charge density.

We now show that the solution of Eq.~(\ref{eom}) is a manageable problem, and that the resulting solution
obeys the Luttinger relation on the Fermi surface volume of a Fermi liquid. 

We make a simple choice of metric of a
truncated AdS$_4$ space with unit radius
\beq
ds^2 = \frac{1}{z^2} \left( dz^2 + dt^2 + dx^2 + dy^2 \right)
\eeq
which is terminated at a hard wall $z=z_m$; the same choice is made in a popular model of confinement
in QCD \cite{ssqcd,igor}. We will specify boundary conditions of the fields at $z=z_m$ below.

It is now useful to write down the explicit form of the Dirac equation, whose solution determines the determinant
in Eq.~(\ref{free}). We follow Refs.~\onlinecite{hong1,trivedi} and reduce the 4-component Dirac equation to a
2-component equation; the equation with spatial wavevector $k$ along the $x$ direction and energy $E_\ell (k)$
is obtained from Eqs.~(3.11-3.13) in Ref.~\onlinecite{trivedi} to be
\beq
\left(  i \sigma^y  \frac{d}{dz} -  \sigma^x \frac{m}{z}  - k \sigma^z   - q \Phi (z) \right) \chi_{\ell,k} (z) = E_\ell (k) \, \chi_{\ell,k} (z),
\label{diracz}
\eeq
where $\chi$ is a two-component spinor, $\sigma$ are Pauli matrices acting on the spinor space, and $\ell$ is a discrete label
for the energy eigenvalues. These eigenvalues are quantized by our choice of boundary conditions. For $z \rightarrow 0$, we choose
the usual solution in the absence of sources \cite{denef}:
\beq
\chi (z \rightarrow 0) \sim z^m .
\eeq
For the boundary at $z=z_m$, we require that the Dirac operator on the left-hand-side of Eq.~(\ref{diracz}) is self-adjoint 
\cite{GN,GP,DHR,HIY}. We normalize our eigenstates so that
\beq
\int_0^{z_m} dz \chi^\dagger_{\ell, k} (z) \chi_{\ell, k} (z) = 1, \label{norm}
\eeq
and then from the $z$ derivative term in the Dirac operator, the self-adjoint condition is
\beq
\chi_1^\dagger (z_m) \sigma^y \chi_2 (z_m) = 0,
\eeq
for any two spinors $\chi_{1,2}$. We can achieve this by taking either their upper or lower components to vanish at $z=z_m$, 
and we will choose the lower component in our numerical solution below. With the Dirac operator self-adjoint, the eigenvalues $E_\ell (k)$
will all be real.

We can now write down simple explicit expressions for the free energy and its functional derivatives. The fermionic contribution
to Eq.~(\ref{free}) is obtained by filling up all the negative energy states at $T=0$, and so the ground state energy density is
\beq
\mathcal{F} = - \frac{1}{2 e^2} \int_0^{z_m} dz  \left( \frac{d \Phi}{d z} \right)^2 + 
\sum_\ell \int \frac{d^2 k}{4 \pi^2} \, E_\ell (k) \, \theta \left(- E_\ell (k) \right).
 \label{freezero}
\eeq
The functional derivatives of $E_\ell (k)$ follow from the Hellman-Feynman theorem. Taking the functional derivative with respect
to $\Phi (z)$, we have from Eq.~(\ref{eom})
\beq
 \frac{1}{e^2} \frac{d^2 \Phi}{d z^2} \
- q \sum_\ell \int \frac{d^2 k}{4 \pi^2} \theta \left(- E_\ell (k) \right) \chi_{\ell,k}^\dagger (z) \chi_{\ell,k} (z)  = 0. \label{gaussz}
\eeq
This is nothing but Gauss's law for the bulk electric field, with the second-term equal to the bulk charge density;
note that the latter is not simply a local function of $\Phi (z)$, unlike previous approaches \cite{polchinski,sean1,sean2,sean3,eric,hong4}.
We can now integrate Eq.~(\ref{gaussz}) over $z$ from $z=0$ to $z=z_m$, and obtain using Eqs.~(\ref{surface}) and (\ref{norm})
\beq 
\langle \mathcal{Q} \rangle + \left. \frac{1}{e^2} \frac{d \Phi}{d z} \right|_{z=z_m} = 
q \sum_\ell \int \frac{d^2 k}{4 \pi^2} \theta \left(- E_\ell (k) \right) 
\eeq
This equation is our Luttinger relation: the right-hand-side is the total area enclosed by the Fermi surface,
and this equals $\langle \mathcal{Q} \rangle$ modulo any electric flux leaking out from the IR boundary 
at $z=z_m$ \cite{ssffl,seanr,liza}, as could happen in a non-Fermi liquid phase. We want a Fermi liquid
solution here, and so we have our final boundary condition for the electrochemical potential
\beq
\left. \frac{d \Phi}{d z} \right|_{z=z_m} = 0. \label{bounde}
\eeq
With this boundary condition, our solution obeys the conventional Luttinger theorem for a Fermi liquid.

\subsection{Numerical solution}

This solution will describe the solution of Eqns.~(\ref{diracz}) and (\ref{gaussz}), subject to the boundary conditions described above.

First, at $\mu=0$, the Dirac equation is solved by Bessel functions
\beq
\chi_{\ell k} (z) \propto \sqrt{z} \left( \begin{array}{c} 
- \displaystyle \frac{M_\ell }{(k+E_\ell (k))} J_{m+1/2} (M_\ell z) 
\\
J_{m-1/2} (M_\ell z) 
\end{array} \right). \label{bessel}
\eeq
Using the boundary condition of a vanishing lower component at $z=z_m$, the parameter
$M_\ell$ is determined by the $\ell$'th zero of the Bessel function of order $m-1/2$, $j_{m-1/2,\ell}$, with 
$M_\ell = j_{m-1/2,\ell}/z_m$. The energy eigenvalues are
\beq
E_\ell (k) = \pm \sqrt{k^2 + M_\ell^2}, \label{diracspec}
\eeq
representing Dirac fermions of mass $M_\ell$ in the confined state.
All the negative energy states will be occupied in the Dirac sea, and we measure all charge densities
and electrochemical potentials relative to this Dirac sea. The completeness relation for the wavefunctions 
$\chi_\ell (k)$ implies that the charge density of the filled Dirac sea is $z$-independent;
we explicitly verified the independence on $z$ by a numerical sum over the normalized Eq.~(\ref{bessel}) 
to a large value of $\ell$. This $z$-independence is associated with the $z$-independent measure in Eq.~(\ref{norm}),
and 
confirms our choice of boundary conditions as a 
consistent description of a gapped, confining state with the spectrum in Eq.~(\ref{diracspec}).
This (infinite) constant charge density is implicitly subtracted from all our expressions.

Turning on a chemical potential, we see no change in the solution as long as $|\mu| < M_1$, the smallest
Dirac mass. So the Dirac vacuum is gapped and incompressible. 
As $\mu$ crosses $M_1$, an additional band of previously positive energy states will be occupied,
and we will obtain a gapless compressible Fermi liquid. To describe this state,
we write
\beq
\chi_{\ell,k} (z) = C z^m \left( 
\begin{array}{c} f_2 (z) \\  f_1 (z) 
\end{array}
\right) \label{c3}
\eeq
where $C$ is a constant set by the normalization in Eq.~(\ref{norm}).
The wavefunction components obey the boundary conditions 
\bea 
f_1 (0) &=& 1 \nn
f_1 (z_m) &=& 0 \nn
f_2 (z \rightarrow 0) &=& -  \frac{(E_\ell (k) + q \Phi (z) - k)}{(2m+1)} z , 
\label{eig1}
\eea
and the differential equations
\bea
\frac{df_1}{dz} &=&  \left( E_\ell (k) + q \Phi (z) + k \right) f_2 (z) \nn
\frac{df_2}{dz} &=& -  (E_\ell (k) + q \Phi (z) - k)  f_1 (z) - \frac{2m}{z} f_2 (z) .
\label{eig2}
\eea
We made an initial (arbitrary) choice of $\Phi (z)$, and numerically solved the eigenvalue
problem defined by Eqs.~(\ref{eig1}) and (\ref{eig2}). 
\begin{figure}[tbp]
  \centering
  \includegraphics[width=5in]{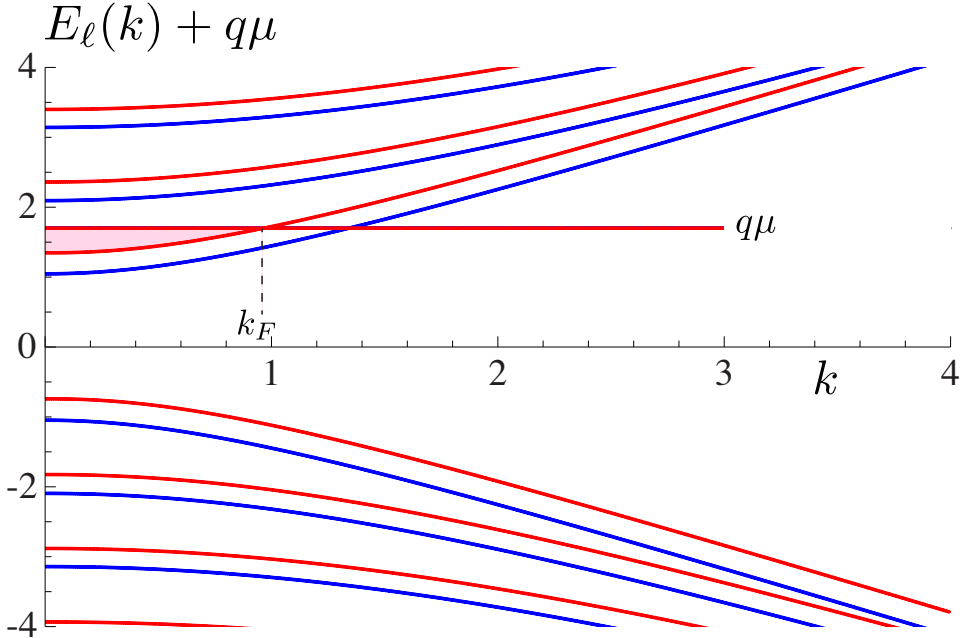} 
  \caption{Dispersion spectrum of the fermions. The blue lines correspond to the spectrum Eq.~(\ref{diracspec},
  at $m=1$, $z_m = 3$, and $\mu=0$. The red lines are at $q\mu=1.7$ and $q^2 e^2 =3$. The horizontal red line
  is at $q\mu$, and the shaded region shows the additional states filled by the chemical potential. Note that the two sets
  of band dispersions are offset even at large momenta and energies: this arises from a Hartree shift in the energies
  due to the added density of particles. The wavefunctions of the states at large momenta are not modified by this shift.
  At smaller momenta, both the $k$ dependence and wavefunctions are different between the two sets.
  }
  \label{fig:disp}
\end{figure}
After determining the eigenvalues and
eigenfunctions, we computed the electric field, $\mathcal{E}(z)$, by integration
\beq
\mathcal{E} (z) = q e^2 \sum_\ell \int \frac{d^2 k}{4 \pi^2} \theta \left(- E_\ell (k) \right) \int_z^{z_m} dz \, \chi_{\ell,k}^\dagger (z) \chi_{\ell,k} (z) . \label{c1}
\eeq
Determination of the electrochemical potential required one more integration
\beq
\Phi (z) =  \mu - \int_0^{z} dz \, \mathcal{E}(z) \label{c2}
\eeq
It can be verified that the boundary conditions in Eqs.~(\ref{eom}) and (\ref{bounde}) are satisfied.
After this determination of $\Phi (z)$, we returned to the solution of Eqs.~(\ref{eig1}) and (\ref{eig2}),
and iterated the procedure until the solution converged. We found little difficulty in the iteration,
and the solution typically converged accurately after $\sim 30$ iterations.

\begin{figure}[tbp]
  \centering
  \includegraphics[width=3in]{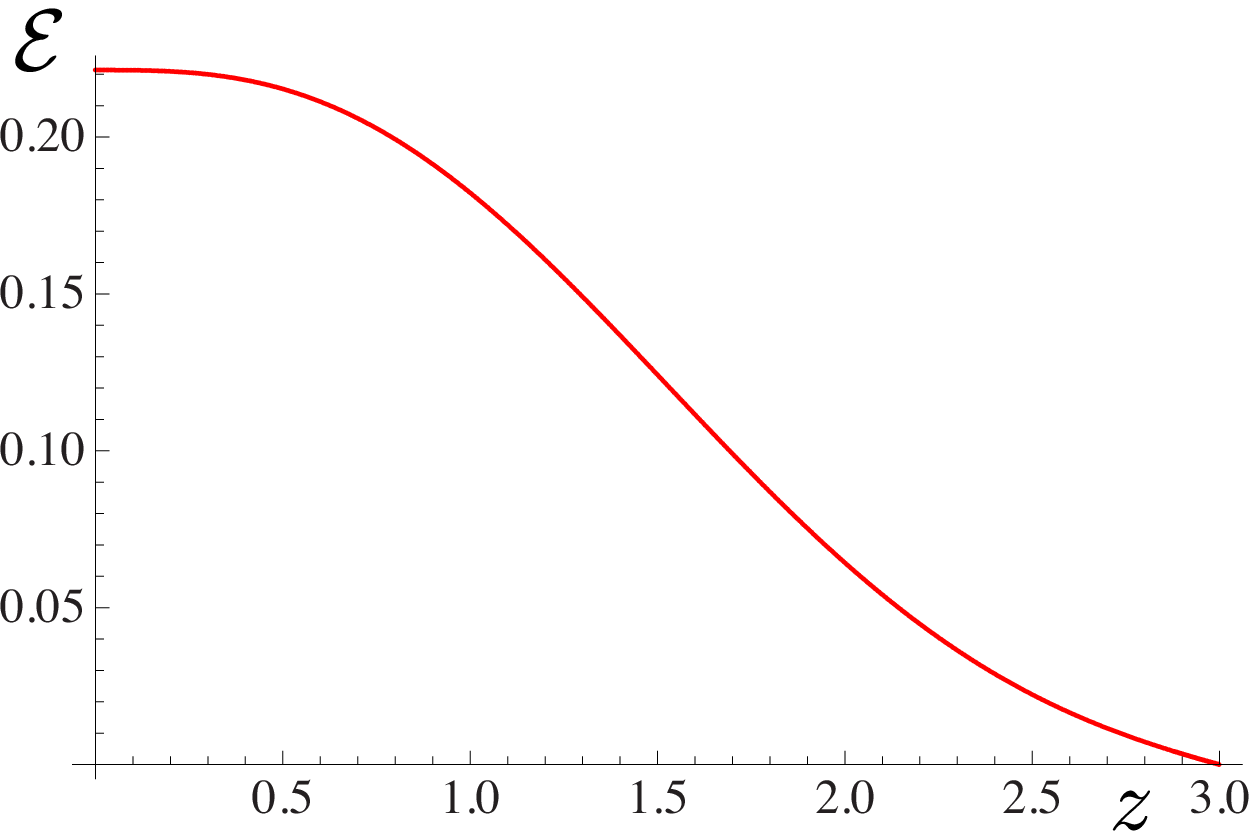} \\ \includegraphics[width=3in]{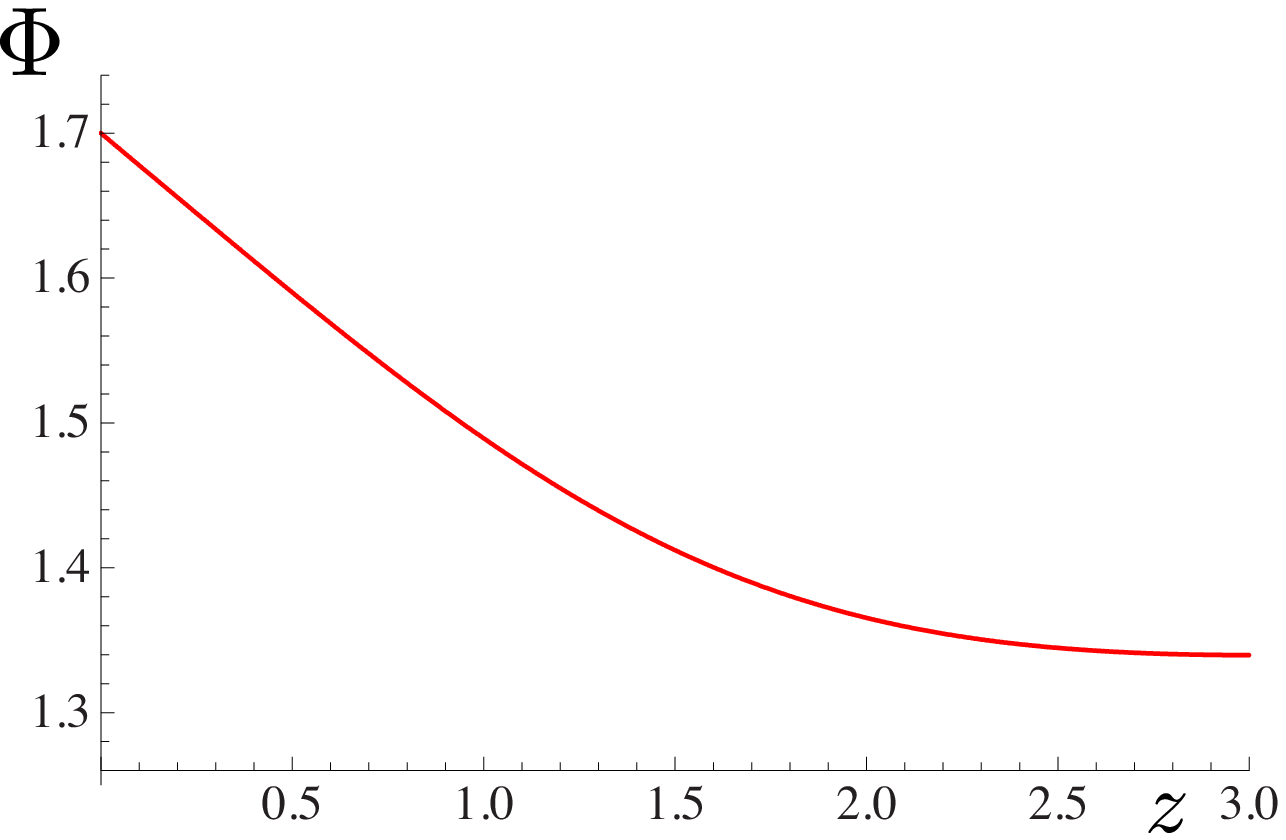}
  \caption{Electric field and electrochemical potential for a non-zero fermion density. Parameters as in Fig.~\ref{fig:disp}.}
  \label{fig:elec}
\end{figure}
\begin{figure}[htbp]
  \centering
  \includegraphics[width=3in]{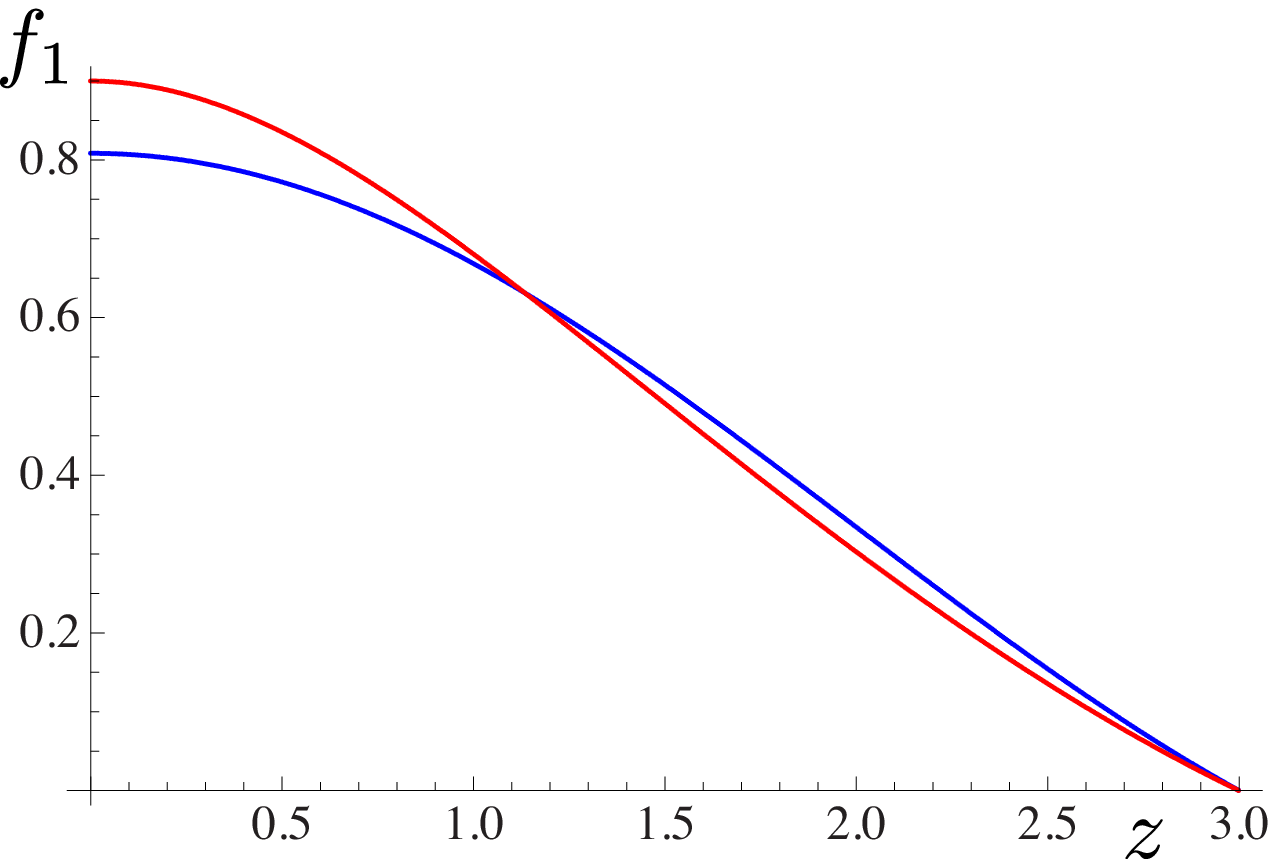} \\
    \includegraphics[width=3in]{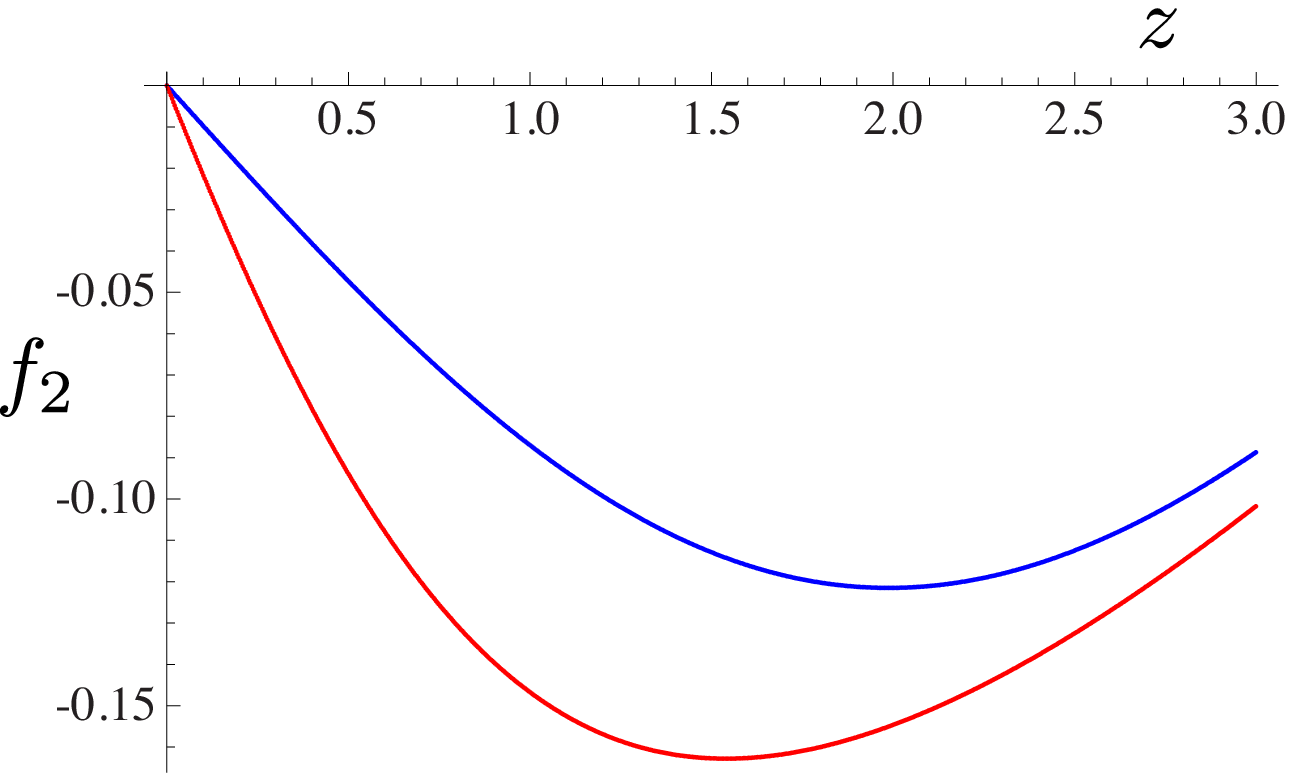} 
  \caption{Wavefunctions at $k=k_F$ with parameters as in Fig.~\ref{fig:disp}. The blue lines show the zero density state in 
  Eq.~(\ref{bessel}), plotted at the wavevector with eigenenergy 1.7. The red lines are the state at the Fermi level
  in the finite density fermion system with $\mu=1.7$. The differences in the two lines are the consequences
  of the self-consistent electric field generated by the fermion density.}
  \label{fig:f1}
\end{figure}
We show our results for the dispersion $E_\ell (k)$ in Fig.~\ref{fig:disp}, both with and without a chemical potential.
The electric field and electrochemical potential computed from Eq.~(\ref{c1},\ref{c2}) are shown in Fig.~\ref{fig:elec},
and the wavefunctions in Fig.~\ref{fig:f1}.
In computing the electric field, we only included the single $\ell$ value associated with the shaded states in Fig.~\ref{fig:disp},
in the summation in Eq.~(\ref{c1}).
Strictly speaking, we should also account for the fact that the states in the negative Dirac sea have also had
their wavefunctions modified by the change in fermion density, and so will contribute to the electric field. However, we expect
such contributions to be suppressed by the Dirac mass gap, and neglected them for simplicity.
The more significant question is whether such corrections can be divergent after summing over the
infinite number of occupied negative energy states, but this is difficult to answer reliably using numerics.
From the perspective of the boundary theory, we expect that all ultraviolet divergencies are associated
with the conformal field theory, and introducing the IR scale $\mu$ does not introduce new UV divergencies; consequently
we do not expect divergencies associated with the electric field.

\section{Beyond mean-field theory}
\label{sec:beyond}

We now extend our results to include full quantum fluctuations of the quantum-electrodynamic theory
in Eq.~(\ref{qed}). In principle, our arguments also allow for fluctuations of other fields, including the 
metric (suitably regularized). We will show that the state obtained in Section~\ref{sec:mft} is a Landau Fermi liquid.

The Gauss's Law result in Eq.~(\ref{gaussz}) has a generalization in the full quantum theory. We write the Ward identity
associated with the equation of motion of $\Phi$ as
\beq
\frac{1}{e^2} \frac{d^2 \langle \Phi \rangle}{dz^2} + q  \int \frac{d\omega}{\pi} \int \frac{d^2 k}{4 \pi^2} \mbox{Im} G^R (\omega, k, z, z) n_F (\omega)
 = 0
\label{gaussf}
\eeq
where $G^R (\omega, k, z, z')$ is the exact bulk-to-bulk retarded Green's function for the fermion field, and $n_F (\omega) = (e^{\omega/T}+1)^{-1}$
is the Fermi distribution function. 
In our mean-field theory,
this Green's function has the spectral representation
\beq
G^{R}_0 (\omega, k, z, z') = \sum_\ell \frac{ \chi_{\ell,k}^\dagger (z) \chi_{\ell,k} (z')}{\omega - E_{\ell} (k) + i 0^+}, \label{gr}
\eeq
and inserting this into Eq.~(\ref{gaussf}), we revert to Eq.~(\ref{gaussz}) at $T=0$. The expression in Eq.~(\ref{gr}) can be used to obtain
response functions of the boundary theory by sending $z, z' \rightarrow 0$ \cite{klebwitt,iqballiu}.

Now our central point is that we can treat the bulk 
$G^R$ as the Green's function of an inhomogenous 3-dimensional Fermi liquid with
short-range interactions. This 3-dimensional system is confined in a potential along with $z$ direction, and translationally invariant
under the infinite $x$ and $y$ directions. The boundary conditions along the $z$ direction ensure that the transverse
components of the gauge field are gapped \cite{ssqcd}, and so there is no long-range interaction along the $x$ and $y$ directions.
Thus, we can apply the conventional methods of many body theory to this inhomogenous system. These methods imply that the
trace of the Green's function along the inhomogenous direction must obey a Luttinger theorem on the volumes of
the Fermi surfaces associated with motion along the $x$ and $y$ directions. In other words, we can express the $z$ motion in terms
of a discrete set of ``bands'' with label $\ell$, and then the system looks very similar to a two-dimensional multiband solid state
system. The Luttinger theorem applies to the sum over all bands, or equivalents, to the trace along the $z$ co-ordinate. So
we conclude that at $T=0$
\beq
- \int_0^{z_m} dz \int \frac{d\omega}{\pi} \int \frac{d^2 k}{4 \pi^2} \mbox{Im} G^R (\omega, k, z, z) \theta (-\omega) = 
\frac{\mbox{Areas enclosed by Fermi surfaces}}{4 \pi^2}.
\eeq
The Luttinger result for the boundary theory now follows from application of Eq.~(\ref{surface}) to 
the $z$ integral of Eq.~(\ref{gaussf}), just as in the mean-field theory. 

Similar arguments, making the analogy to a multi-band two-dimensional Fermi liquid with short-range interactions, 
imply that this system realizes all other properties of a Landau Fermi liquid. In particular, transverse gauge modes
which are gapped and discrete at Gaussian order in the holographic theory \cite{ssqcd}, will acquire Landau damping from
the Fermi surface excitations at higher orders, and so their spectrum will eventually become continuous. Similar comments
apply to the spectrum of gravitons and other bosonic modes. 

\section{Conclusions}
\label{sec:conc}

We have presented a simple model for the holographic description of a Fermi liquid.
We begin with a conformal field theory in 2+1 dimensions \cite{tasi,semenoff}, which has a holographic description
by a theory on AdS$_4$. Then we change its carrier density by
applying a chemical potential $\mu$. Using arguments that the Fermi liquid must be a confining state \cite{liza},
we assumed that the AdS$_4$ geometry was truncated by confinement at a distance
$z_m$ along the holographic direction. The resulting state is described by 2 energy scales, $\mu$ and $1/z_m$.
We presented a theory for the crossover across these energy scales, and showed that a Fermi liquid is obtained
at the lowest energy scales. The bulk gauge field was determined self-consistently, and this was closely
linked to consistency with Luttinger's theorem.

The main shortcoming of our analysis was that the confinement was essentially put in by hand, and not obtained
by a self-consistent determination of the metric. In principle, confinement can be a direct consequence of introducing
a non-zero $\mu$ on the conformal field theory. In this case, we expect the value of the confinement scale, $1/z_m$, 
to determined by $\mu$, and to vanish as $\mu \rightarrow 0$; such a situation would be analogous to `deconfined criticality' \cite{dcp1,dcp2}.
In particular, the scaling properties of the conformal field theory imply that $z_m = \mathcal{C}/\mu$, where $\mathcal{C}$
is a universal constant characteristic of the conformal field theory. We leave the possible realization of such a scenario,
and computation of $\mathcal{C}$, as important problems for future research.

\acknowledgements
I thank S.~Hartnoll and L.~Huijse for valuable discussions and collaborations on related projects.
I am grateful to M.~\v{C}ubrovi\'{c}, 
T.~Faulkner, C.~Herzog, D.~Hofman, A.~Karch, E.~Kiritsis, H.~Liu, J.~McGreevy, R.~Myers, P.~Phillips, K.~Schalm, D.~Son, L.~Thorlacius, 
S.~Trivedi, and J. Zaanen for helpful discussions.
This research was supported by the National Science Foundation under grant DMR-0757145, by a MURI grant from AFOSR.


\begin{thebibliography}{}

\bibitem{nernst} 
  S.~A.~Hartnoll, P.~K.~Kovtun, M.~M\"uller, and S.~Sachdev,
  ``Theory of the Nernst effect near quantum phase transitions in condensed matter, and in dyonic black holes,''
  Phys.\ Rev.\  {\bf B76}, 144502 (2007)
  [arXiv:0706.3215 [cond-mat.str-el]].

\bibitem{sslee0}
  S.-S.~Lee,
  ``A Non-Fermi Liquid from a Charged Black Hole: A Critical Fermi Ball,''
  Phys.\ Rev.\  D {\bf 79}, 086006 (2009)
  [arXiv:0809.3402 [hep-th]].
  
\bibitem{denef0}
F.~Denef and S.~A.~Hartnoll,
  ``Landscape of superconducting membranes,''
  Phys.\ Rev.\ D {\bf 79}, 126008 (2009)
  [arXiv:0901.1160 [hep-th]].  

\bibitem{hong0}
  H.~Liu, J.~McGreevy, and D.~Vegh,
  ``Non-Fermi liquids from holography,''
  arXiv:0903.2477 [hep-th].
  
  \bibitem{zaanen1}
  M.~\v{C}ubrovi\'{c}, J.~Zaanen, and K.~Schalm,
  ``String Theory, Quantum Phase Transitions and the Emergent Fermi-Liquid,''
  Science {\bf 325}, 439 (2009).
  [arXiv:0904.1993 [hep-th]].

\bibitem{hong1} T.~Faulkner, H.~Liu, J.~McGreevy, and D.~Vegh,
  ``Emergent quantum criticality, Fermi surfaces, and AdS(2),''
  arXiv:0907.2694 [hep-th].
 
\bibitem{denef}
  F.~Denef, S.~A.~Hartnoll, and S.~Sachdev,
  ``Quantum oscillations and black hole ringing,''
  Phys.\ Rev.\ D {\bf 80}, 126016 (2009)
  [arXiv:0908.1788 [hep-th]].
     
\bibitem{faulkner} T.~Faulkner and  J.~Polchinski,
  ``Semi-Holographic Fermi Liquids,''
  arXiv:1001.5049 [hep-th].

\bibitem{polchinski}  S.~A.~Hartnoll, J.~Polchinski, E.~Silverstein, and D.~Tong,
  ``Towards strange metallic holography,''
  JHEP {\bf 1004}, 120 (2010)
  [arXiv:0912.1061 [hep-th]].
  
\bibitem{gubserrocha}    S.~S.~Gubser and F.~D.~Rocha,
  ``Peculiar properties of a charged dilatonic black hole in AdS$_5$,''
  Phys.\ Rev.\  D {\bf 81}, 046001 (2010)
  [arXiv:0911.2898 [hep-th]].

\bibitem{hong2}
  T.~Faulkner, N.~Iqbal, H.~Liu, J.~McGreevy, and D.~Vegh,
  ``Strange metal transport realized by gauge/gravity duality,''
  Science {\bf 329}, 1043 (2010).
  
\bibitem{kiritsis}   C.~Charmousis, B.~Gouteraux, B.~S.~Kim, E.~Kiritsis, and R.~Meyer,
  ``Effective Holographic Theories for low-temperature condensed matter systems,''
  arXiv:1005.4690 [hep-th].  

\bibitem{sean1}  S.~A.~Hartnoll and A.~Tavanfar,
  ``Electron stars for holographic metallic criticality,''
  arXiv:1008.2828 [hep-th].

\bibitem{sean2}   S.~A.~Hartnoll, D.~M.~Hofman, and A.~Tavanfar,
  ``Holographically smeared Fermi surface: Quantum oscillations and Luttinger count in electron stars,''
  arXiv:1011.2502 [hep-th].

\bibitem{sean3}   S.~A.~Hartnoll, D.~M.~Hofman, and D.~Vegh,
  ``Stellar spectroscopy: Fermions and holographic Lifshitz criticality,''
  arXiv:1105.3197 [hep-th].
  
\bibitem{seanr}   S.~A.~Hartnoll,
  ``Horizons, holography and condensed matter,''
  arXiv:1106.4324 [hep-th].  
  
\bibitem{larus1}   E.~J.~Brynjolfsson, U.~H.~Danielsson, L.~Thorlacius, and T.~Zingg,
  ``Black Hole Thermodynamics and Heavy Fermion Metals,''
  JHEP {\bf 1008}, 027 (2010)
  [arXiv:1003.5361 [hep-th]].  
  
\bibitem{larus2}   V.~G.~M.~Puletti, S.~Nowling, L.~Thorlacius, and T.~Zingg,
  ``Holographic metals at finite temperature,''
  JHEP {\bf 1101}, 117 (2011)
  [arXiv:1011.6261 [hep-th]].  

\bibitem{eric} X.~Arsiwalla, J.~de Boer, K.~Papadodimas, and E.~Verlinde,
  ``Degenerate Stars and Gravitational Collapse in AdS/CFT,''
  arXiv:1010.5784 [hep-th].

\bibitem{kachru2}  
    K.~Goldstein, S.~Kachru, S.~Prakash, and S.~P.~Trivedi,
  ``Holography of Charged Dilaton Black Holes,''
  JHEP {\bf 1008}, 078 (2010)
  [arXiv:0911.3586 [hep-th]].

\bibitem{kachru3}  
    K.~Goldstein, N.~Iizuka, S.~Kachru, S.~Prakash, S.~P.~Trivedi, and A.~Westphal,
  ``Holography of Dyonic Dilaton Black Branes,''
  JHEP {\bf 1010}, 027 (2010)
  [arXiv:1007.2490 [hep-th]].
  
\bibitem{kachru4}   K.~Jensen, S.~Kachru, A.~Karch, J.~Polchinski, and E.~Silverstein,
  ``Towards a holographic marginal Fermi liquid,''
  arXiv:1105.1772 [hep-th].

\bibitem{trivedi}   N.~Iizuka, N.~Kundu, P.~Narayan, and S.~P.~Trivedi,
  ``Holographic Fermi and Non-Fermi Liquids with Transitions in Dilaton Gravity,''
  arXiv:1105.1162 [hep-th].

\bibitem{zaanen2}
  M.~\v{C}ubrovi\'{c}, J.~Zaanen, and K.~Schalm,
  ``Constructing the AdS dual of a Fermi liquid: AdS Black holes with Dirac hair,''
  arXiv:1012.5681 [hep-th].
  
\bibitem{ssffl}   S.~Sachdev,
  ``Holographic metals and the fractionalized Fermi liquid,''
  Phys.\ Rev.\ Lett.\  {\bf 105}, 151602 (2010)
  [arXiv:1006.3794 [hep-th]].

\bibitem{mcphys} J.~McGreevy, ``In pursuit of a nameless metal,'' Physics {\bf 3}, 83 (2010).

\bibitem{liza} L.~Huijse and S.~Sachdev, ``Fermi surfaces and gauge-gravity duality,''
Phys. Rev. D {\bf 84}, 026001 (2011) [arXiv:1104.5022 [hep-th]].  

\bibitem{leiden}   M.~\v{C}ubrovi\'{c}, Y.~Liu, K.~Schalm, Y.-W.~Sun, and J.~Zaanen,
  ``Spectral probes of the holographic Fermi groundstate: dialing between the electron star and AdS Dirac hair,''
  arXiv:1106.1798 [hep-th].
  
\bibitem{hong4}   N.~Iqbal, H.~Liu, and M.~Mezei,
  ``Semi-local quantum liquids,''
  arXiv:1105.4621 [hep-th].

\bibitem{pp} M.~Edalati, K.~W.~Lo, and P.~W.~Phillips,
  ``Neutral Order Parameters in Metallic Criticality in $d=2+1$ from a Hairy Electron Star,''
  arXiv:1106.3139 [hep-th].  
  
\bibitem{waldram} J.~P.~Gauntlett, J.~Sonner, and D.~Waldram,  
``The spectral function of the supersymmetry current (I),''
arXiv:1106.4694 [hep-th].

\bibitem{yarom} R.~Belliard, S.~S.~Gubser, and Amos Yarom,
``Absence of a Fermi surface in classical minimal four-dimensional gauged supergravity,'' 
arXiv:1106.6030 [hep-th].


\bibitem{tasi} S.~Sachdev, ``The landscape of the Hubbard model,'' Section V
in {\em TASI 2010, String Theory and Its Applications: From meV to the Planck Scale,\/}
M.~Dine, T.~Banks, and S.~Sachdev Eds., World Scientific, Singapore (2011),
arXiv:1012.0299 [hep-th].

\bibitem{semenoff}   J.~L.~Davis, H.~Omid, and G.~W.~Semenoff,
  ``Holographic Fermionic Fixed Points in $d=3$,''
  arXiv:1107.4397 [hep-th].
  
\bibitem{witten} E. Witten, ``Anti-de Sitter space, thermal phase transition, and confinement in gauge theories,'' 
Adv. Theor. Math. Phys. {\bf 2}, 505 (1998) [arXiv:hep-th/9803131].

\bibitem{meyer} G.~T.~Horowitz and R.~C.~Myers, ``The AdS/CFT Correspondence and a 
New Positive Energy Conjecture for General Relativity," Phys. Rev. D {\bf 59}, 026005 (1998) 
[arXiv:hep-th/9808079].

\bibitem{tadashi}   T.~Nishioka, S.~Ryu, and T.~Takayanagi,
  ``Holographic Superconductor/Insulator Transition at Zero Temperature,''
  JHEP {\bf 1003}, 131 (2010)
  [arXiv:0911.0962 [hep-th]].

\bibitem{gary} G.~T.~Horowitz and B.~Way,
  ``Complete Phase Diagrams for a Holographic Superconductor/Insulator System,''
  JHEP {\bf 1011}, 011 (2010)
  [arXiv:1007.3714 [hep-th]].

\bibitem{igor}  I.~R.~Klebanov and M.~J.~Strassler,
  ``Supergravity and a confining gauge theory: Duality cascades and $\chi$SB resolution of naked singularities,''
  JHEP {\bf 0008}, 052 (2000)
  [arXiv:hep-th/0007191].

\bibitem{ssqcd}   J.~Erlich, E.~Katz, D.~T.~Son, and M.~A.~Stephanov,
  ``QCD and a holographic model of hadrons,''
  Phys.\ Rev.\ Lett.\  {\bf 95}, 261602 (2005)
  [arXiv:hep-ph/0501128].
  
\bibitem{soft1}   A.~Karch, E.~Katz, D.~T.~Son, and M.~A.~Stephanov,
  ``Linear confinement and AdS/QCD,''
  Phys.\ Rev.\ D\  {\bf 74}, 015005 (2006)
  [arXiv:hep-ph/0602229].  
  
\bibitem{soft2}   U.~Gursoy and E.~Kiritsis,
  ``Exploring improved holographic theories for QCD: Part I,''
  JHEP {\bf 0802}, 032 (2008)
  [arXiv:0707.1324 [hep-th]].

\bibitem{soft3}   U.~Gursoy, E.~Kiritsis, L.~Mazzanti, and F.~Nitti,
  ``Holography and Thermodynamics of 5D Dilaton-gravity,''
  JHEP {\bf 0905}, 033 (2009)
  [arXiv:0812.0792 [hep-th]].

\bibitem{soft4}    S.~S.~Gubser and A.~Nellore,
  ``Mimicking the QCD equation of state with a dual black hole,''
  Phys.\ Rev.\  D\ {\bf 78}, 086007 (2008)
  [arXiv:0804.0434 [hep-th]].
  
\bibitem{GN} Y.~Grossman and M.~Neubert,
  ``Neutrino masses and mixings in nonfactorizable geometry,''
  Phys.\ Lett.\  B {\bf 474}, 361 (2000)
  [arXiv:hep-ph/9912408].
  
\bibitem{GP} T.~Gherghetta and A.~Pomarol,
  ``Bulk fields and supersymmetry in a slice of AdS,''
  Nucl.\ Phys.\ B {\bf 586}, 141 (2000)
  [arXiv:hep-ph/0003129].
  
\bibitem{DHR}  H.~Davoudiasl, J.~L.~Hewett, and T.~G.~Rizzo,
  ``Experimental probes of localized gravity: On and off the wall,''
  Phys.\ Rev.\ D {\bf 63}, 075004 (2001)
  [arXiv:hep-ph/0006041].

\bibitem{HIY}  D.~K.~Hong, T.~Inami, and H.~-U.~Yee,
  ``Baryons in AdS/QCD,''
  Phys.\ Lett.\  B {\bf 646}, 165 (2007)
  [arXiv:hep-ph/0609270].
  
\bibitem{klebwitt}   I.~R.~Klebanov and E.~Witten,
  ``AdS / CFT correspondence and symmetry breaking,''
  Nucl.\ Phys.\  {\bf B556}, 89-114 (1999)
  [arXiv:hep-th/9905104].

\bibitem{iqballiu}  N.~Iqbal and H.~Liu,
  ``Real-time response in AdS/CFT with application to spinors,''
  Fortsch.\ Phys.\  {\bf 57}, 367 (2009)
  [arXiv:0903.2596 [hep-th]].
  
\bibitem{dcp1} T.~Senthil, A.~Vishwanath, L.~Balents, S.~Sachdev, and
M.~P.~A.~Fisher, ``Deconfined quantum critical points,'' Science
{\bf 303}, 1490 (2004) [arXiv:cond-mat/0311326].
  
\bibitem{dcp2} S. Sachdev and X. Yin,
``Quantum phase transitions beyond the Landau-Ginzburg paradigm and supersymmetry,''  
Annals of Physics {\bf 325}, 2 (2010) [arXiv:0808.0191 [hep-th]].
  
\end{thebibliography}
\end{document}